\documentclass[
	twocolumn]
	{biophys-new}
\usepackage{helvet,times}
\usepackage{bm,textcomp}

\usepackage{csquotes}
\usepackage{graphicx} 
\usepackage{comment}
\usepackage{color}
\usepackage{hyperref}
\usepackage[caption=false,]{subfig}
\usepackage{enumerate}
\usepackage{float}
\usepackage{amsmath}
\usepackage{amssymb}

\newcommand{\mum}{\mu \textnormal{m}}

\newcommand{\nM}{\textnormal{nM}}
\renewcommand{\sec}{\textnormal{s}}
\newcommand{\Prob}{\textnormal{Prob}}

\title{Crowding and pausing strongly affect dynamics of kinesin-1 motors along microtubules}
\runningtitle{Crowding and pausing of kinesin-1} 

\author[1]{Matthias Rank}
\author[1,*]{Erwin Frey}
\runningauthor{Matthias Rank and Erwin Frey} 

\affil[1]{Arnold-Sommerfeld-Center for Theoretical Physics and Center for NanoScience, Ludwig-Maximilians-Universit\"{a}t M\"{u}nchen, Theresienstra{\ss}e 37, 80333 M\"{u}nchen, Germany}

\corrauthor[*]{frey@lmu.de}

\papertype{Article}

\begin{document}  

\begin{frontmatter}

\begin{abstract}
	{
	Molecular motors of the kinesin-1 family move in a directed and processive fashion along microtubules (MTs).
	It is generally accepted that steric hindrance of motors leads to crowding effects; however, little is known about the specific interactions involved.
	We employ an agent-based lattice gas model to study the impact of interactions which enhance the detachment of motors from crowded filaments on their collective dynamics.
	The predictions of our model quantitatively agree with the experimentally observed concentration dependence of key motor characteristics including their run length, dwell time, velocity, and landing rate. 
	From the anomalous stepping statistics of individual motors which exhibit relatively long pauses we infer that kinesin-1 motors sometimes lapse into an inactive state. 
	Hereby, the formation of traffic jams amplifies the impact of single inactive motors and leads to a crowding dependence of the frequencies and durations of the resulting periods of no or slow motion.
	We interpret these findings and conclude that kinesin-1 spends a significant fraction of its stepping cycle in a weakly bound state in which only one of its heads is bound to the MT.
}
{}
{}
\end{abstract}

\end{frontmatter}

\section*{Introduction}
	The collective motion of molecular motors on microtubules (MTs) and their interactions with each other are highly complex processes that underlie important intracellular functions.
	For example, motors of the kinesin-8 family use MTs as molecular tracks along which they perform directed transport~\cite{Gupta2006,Varga2006}.
	Having arrived at the MT end, these motors influence the depolymerisation dynamics at this point, and thus have an effect on MT length~\cite{Varga2006,Varga2009,Melbinger2012,Rank2018} and spindle size~\cite{Goshima2005,Stumpff2008}, properties whose tight regulation is crucial for the normal operation of a cell~\cite{Jordan2004}.

	Kinesin-1 was the first kinesin to be discovered~\cite{Vale1985}, and it is arguably the motor which has been studied in greatest detail. 
	Kinesin-1 is a versatile cargo transporter~\cite{Hirokawa2005} which uses its two heads~\cite{Hirokawa1998} to processively walk towards the plus-end of a MT.
	In the crowded environment of a typical cell, molecular motors and MT-associated proteins~\cite{Akhmanova2008} compete for a limited number of binding sites on the MTs. 
	As a consequence, \enquote{traffic jams} consisting of molecular motors may develop on (parts of) the MT~\cite{Leduc2012,Subramanian2013}.

	A central question is how motors interact with each other in crowded situations like this, and how motors affect each other's ability to bind to and detach from MTs. 
	Several studies have reported (apparently) conflicting results relating to these issues: Thus, Vilfan et al.~\cite{Vilfan2001a} observed that kinesin motors primarily bind near other motors. Similarly, Muto et al.~\cite{Muto2005} observed long-range cooperative binding, and Roos et al.~\cite{Roos2008} discovered that the dwell time of motors increases when they are in the proximity of other motors on the MT.
	In contrast, Leduc et al.~\cite{Leduc2012} found a reduction in the dwell time of kinesin-8 motors on crowded filaments, in agreement with \emph{in vitro} measurements of kinesin-1 carried out by Telley et al.~\cite{Telley2009}.

	How can these findings be reconciled? Firstly, we note that interactions may differ depending on whether motors are mobile~\cite{Telley2009,Leduc2012} or have been immobilized by genetic engineering~\cite{Vilfan2001a,Roos2008}: It appears that an increased dwell time of motors on the MT or cooperative attachment to a MT is primarily found for immobile motors, while mobile motors experience no, or at least less attractive interactions.
	A second differentiator of these studies was pointed out by Telley et al.~\cite{Telley2009} who found that the label used to visualise motors by fluorescence microscopy can be crucial.
	In particular, when these authors failed to reproduce their own earlier results~\cite{Seitz2006a} for the crowding behaviour of kinesin-1 using a different label, they concluded that extensive labelling or the use of large labels may lead to non-specific interactions between motors.
Therefore, attractive potentials may develop which hold motors on the MT.

	To minimise these potential effects, Telley et al. removed parts of kinesin's tail~\cite{Berliner1995}, such that the motor could still walk with wild-type characteristics~\cite{Seitz2006a}, and attached a GFP label to only a small proportion of the motors, leaving the vast majority of kinesin motors unlabelled~\cite{Telley2009}. 
	As a consequence, when they varied the abundance of kinesin, they found that this motor's dwell time was inversely related to its (volume) concentration.
	In our understanding, the situation considered in this study by Telley et al.~\cite{Telley2009} is closest to the behaviour in an actual cell.
Hence, in our theoretical analysis we will mainly compare our results with their data.

	The Totally Asymmetric Simple Exclusion Process with Langmuir Kinetics (TASEP/LK)~\cite{Lipowsky2001,Parmeggiani2003,Parmeggiani2004} is commonly employed to describe the collective dynamics of motors on a MT.
	In this stochastic lattice gas model, motors are described as particles on a one-dimensional lattice (a protofilament of a MT) and step stochastically towards the lattice end.
	This approach has successfully predicted~\cite{Parmeggiani2003,Parmeggiani2004} the existence of traffic jams and domain walls, which were recently observed in experiments~\cite{Leduc2012,Subramanian2013}.
	Several variations of this stochastic process have considered specific properties of motors, such as their longitudinal~\cite{Pierobon2006a} or lateral~\cite{Melbinger2010} extension. 
	Furthermore, additional interactions of motors with each other have been examined~\cite{Celis-Garza2015,Chandel2015,Teimouri2015}.
	Among them are so-called mutually interactive Langmuir kinetics~\cite{Vuijk2015a, Messelink2016, Gupta2015}, where binding and unbinding of monomeric particles are directly influenced by the occupation of the nearest-neighbour binding sites.
	Most of these studies concentrated on fundamental physical properties of the dynamics of motors, such as the different phases of their collective motion.
	Consequently, the impact of motor-motor interactions on experimentally accessible quantities, such as the motor run length, dwell time, velocity or their numbers of landings (initial attachments) on the lattice per unit length and time, was usually not considered.

	In this study, we theoretically examine a model which includes motor-motor interactions and a dimeric driven lattice gas. 
	Our aim is to describe the collective motion of processive molecular motors, such as kinesin-1, along a MT. 
	We find that a simple, motor-induced detachment mechanism suffices to quantitatively account for the experimental measurements reported by Telley et al.~\cite{Telley2009}.
	By developing a mean-field theory, we explore in detail the dependence of motor dwell time, run length, velocity, and landing rate on the volume concentration of kinesin.
	Furthermore, we find that stochastic pausing of motors on the MT is significantly enhanced by crowding and leads to short-lived traffic jams on the MT, thus recovering the long and frequent periods of interrupted motor motion observed in experiments~\cite{Telley2009}.
	By comparing the rates of spontaneous detachment and motor-induced detachment from the MT, we gain insight into the stepping cycle of kinesin-1, and find that this motor spends a significant fraction ($ \,{\sim}\,  22\%$) of its stepping cycle in a weakly bound state.

\section*{\MakeUppercase{Methods}}
\subsection*{Monte Carlo simulations}
	{
	\small
	We simulate our stochastic lattice gas model with Gillespie's algorithm~\cite{Gillespie1977a}, which provides a way of exactly modelling stochastic processes. 
	In the first step, all possible events are collected and statistically weighed with their rates, and an event is randomly chosen out of the resulting vector. 
	Another random number is drawn from an exponential distribution with the total rate (i.e., the sum of the rates of all possible events) as the decay parameter, in order to obtain the update time. 
	Subsequently, all rates are updated and the algorithm starts over.
	In order to account for the long length of MTs compared to the motors' run length (on the order of $100$ steps), periodic boundary conditions were employed on a lattice with $2000$ sites.	
	}

\subsection*{Fitting analytical results to experimental data}
	{
	\small
	For the four sets of quantities measured experimentally~\cite{Telley2009}, namely run length, dwell time, velocity, and landing rate of kinesin motors, analytic equations were obtained, see Eqs.~(\ref{eq_tau})--(\ref{eq_landing}).
	The parameters $\nu$ (hopping rate of motors) and $\omega_D$ (their detachment rate) were obtained from the experimental data~\cite{Telley2009} at low concentrations, as well as the landing rate $\lambda_0$ of normalised concentration of motors to the MT.
	In order to obtain the remaining parameters $\omega_A$ and $\theta$, the analytic results were taken at the concentrations tested in experiments, and the deviations from experimental data were weighed by the experimental standard error~\cite{Telley2009}.
	Subsequently, the sum of the squared weighed errors was taken, and minimized with \emph{Mathematica}'s NMinimize function. 
	In this way, the global fit values $\omega_D$ and $\theta$ are found, see Eq.~(\ref{eq_fitparams}).
	}

\section*{Results}
\subsection*{Model description}

	We wish to analyse the stochastic motion of kinesin-1 motor molecules on MTs. 
	Kinesin-1 is a dimer with two heads~\cite{Hirokawa1998} that can bind to distinct binding sites~\cite{Nogales1998} on two neighbouring tubulin dimers~\cite{Yildiz2004}.
	Powered by the hydrolysis of ATP~\cite{Hua1997}, it moves processively and unidirectionally~\cite{Vale1996} towards the MT's plus-end~\cite{Howard2003} along a protofilament~\cite{Ray1993,Howard1996}.
	It walks hand-over-hand~\cite{Yildiz2004}, which implies that the rear (lagging) head steps over the front (leading) head to the next binding site in order to complete a step.

\begin{figure}[hbt]
	\centering
	\includegraphics[width=.8\columnwidth]{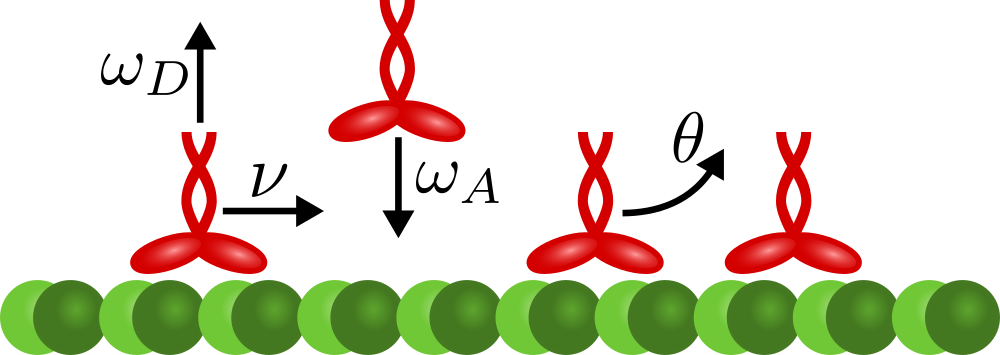}
	\caption{
		\textbf{Lattice gas model for the collective dynamics of kinesin-1 motor proteins moving along a protofilament of a microtubule (MT)}. 
		Motors are modelled as dimers that simultaneously occupy two neighbouring lattice sites, and advance unidirectionally towards the plus-end (right) of a protofilament at a rate $\nu$ (Poisson stepper), if no other motor occupies the next binding site (exclusion process). 
		Kinesin-1 is also assumed to randomly bind to and detach from the protofilament at rates $\omega_A$ and $\omega_D$, respectively. 
		Due to steric exclusion binding is possible only if two adjacent binding sites are empty. 
		In addition to spontaneous detachment with rate $\omega_D$, we also account for facilitated detachment of motors that are immediate neighbours. 
		For specificity, we assume that the dissociation rate of the rear motor, i.e., the motor closer to the minus-end (left) is enhanced by a rate $\theta$.
		\label{fig_model}
	}
\end{figure}

	To describe the collective dynamics of kinesin-1 motors on protofilaments, we employ a one-dimensional lattice gas model as illustrated in Fig.~\ref{fig_model}, where the fluid surrounding the MT can be considered as a homogeneous and constant reservoir of motors with concentration $c$.
	The corresponding mathematical model is based on the Totally Asymmetric Simple Exclusion Process with Langmuir Kinetics (TASEP/LK)~\cite{Parmeggiani2003,Parmeggiani2004}. 
	Here, we extend it to include the dimeric nature of kinesin-1, and consider an additional interaction which accounts for the enhanced detachment of neighbouring motors. 
	To accommodate the extended size of kinesin, and to allow us to adopt simple stepping rules, each motor is described as a rigid particle which simultaneously occupies two sites of a one-dimensional lattice~\cite{Pierobon2006a}.
	The directed motion of motors is modelled as a stepwise stochastic hopping process with rate $\nu$ (Poisson process) towards the plus-end (\emph{totally asymmetric}); stepping is possible only if the target site is not occupied by another motor (\emph{exclusion}).
	In the limit of low coverage of a protofilament, each motor would then move at an average speed $v_0 \,{=}\, \nu a$, where $a \,{=}\, 8.4$ nm~\cite{Hyman1995} is the size of a tubulin heterodimer.
	Motors from the reservoir can attach to the protofilament lattice at rate $\omega_A$ at locations where two adjacent lattice sites are empty. This rate depends on the volume concentration of motors as $\omega_A = \omega_a c$ with a constant $\omega_a$.

	There are two pathways that may lead to the detachment of motors from a protofilament. 
	Firstly, motors may detach spontaneously at a rate $\omega_D$.
	Because this alone cannot explain the decrease in motor dwell time on crowded filaments~\cite{Telley2009}, we secondly assume that motors interact with each other via a process that enhances the detachment rate of motors which are immediate neighbours.
	Specifically, when two motors meet, we assume that the \emph{rear} motor's unbinding rate is enhanced by an additional rate $\theta$; the trailing motor therefore \enquote{bounces off} the leading motor, which is consistent with experiments showing that when kinesin runs into an obstacle on the MT, the motor (and not the obstacle) is likely to detach~\cite{Telley2009,Schneider2015}.
	The opposite case, where the trailing motor \enquote{kicks} the leading motor off the filament, leads to the same phenomena. Alternative scenarios, e.g., enhanced detachment of both motors, have been examined in Ref.~\cite{Vuijk2015a}.

\subsection*{Motor currents and density profiles}
	Two central quantities that characterise the collective transport of kinesin-1 along MTs are the motor density $\rho$ and the motor current $j$. 
	In general, both quantities depend on the position along the MT. 
	At the minus-end, the density is expected to show an initial (approximately) linear increase towards a Langmuir plateau due to an \enquote{antenna effect}~\cite{Parmeggiani2003,Parmeggiani2004,Leduc2012}: 
	This gradient arises from the combined effects of random motor attachment to and detachment from the MT, as well as driven transport along it; the slope of the initial increase is proportional to the attachment rate $\omega_A$.
	Similarly, a density gradient can also be found at the MT's plus-end, in particular for motors which remain bound at this tip for an extended time. 
	Molecular motors with this property include kinesin-8~\cite{Leduc2012} and kinesin-4~\cite{Subramanian2013}; to the best of our knowledge, no such behaviour has been reported for kinesin-1.
	Due to (potential) gradients at the MT's ends, it is generally difficult to determine the full quantitative behaviour of the motor density~\cite{Pierobon2006a,Vuijk2015a}.
	One particular property of kinesin-1, the motor in which we are primarily interested in this study, allows for a significant simplification in this respect: its run length (on the order of $1\,\mu$m~\cite{Telley2009}) is significantly less than the length of typical MTs (usually several $\mu$m~\cite{Jeune-Smith2010}).
	For this reason, the extent of the gradient region is small relative to the MT length, and the density profile is for the most part spatially uniform on the MT for this motor.
	By assuming a very long lattice and/or periodic boundary conditions (see Appendix), one can dispense with the specification of the boundary processes.

\begin{figure}[bt]
	\subfloat{\label{sf_general_rho}}\subfloat{\label{sf_general_j}}
	\includegraphics[width=.99\columnwidth]{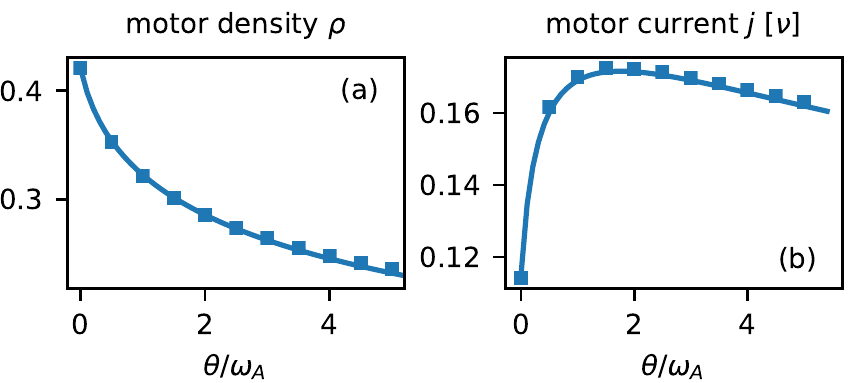} 
	\caption{
		\textbf{Bulk motor density and current}. 
		Symbols show data obtained from stochastic simulations, the lines depict the results of the mean-field analysis, cf. Eqs. (\ref{eq_rhodimer}) and (\ref{eq_j}) for parameters $\omega_A \,{=}\, 0.01 \nu$ and $\omega_D \,{=}\, \omega_A/10$. 
		\protect\subref{sf_general_rho} The interaction-induced unbinding mechanism reduces the motor density $\rho$.  
		\protect\subref{sf_general_j} In contrast, the motor current $j$ reaches a maximum for some finite value of the detachment rate $\theta$. 
	 	\label{fig_general_role_of_theta}
	}
	\centering
\end{figure}

	Figures \ref{sf_general_rho} and \ref{sf_general_j} show the bulk density $\rho$ and current $j$, respectively, as obtained from stochastic simulations using Gillespie's algorithm~\cite{Gillespie1977a}, see Methods.
	We find that the additional detachment of motors facilitated by the interaction between neighbouring motors leads to a monotonic decrease in the bulk density (Fig.~\ref{sf_general_rho}) with increasing rate $\theta$; in the limit $\theta \,{=}\, 0$, we recover previous results~\cite{Pierobon2006a}. 
	Interestingly, the motor current shows non-monotonic behaviour as a function of $\theta$ (Fig.~\ref{sf_general_j}). 
	There is an optimal value of $\theta$ at which the current is maximal.
	This can be understood in terms of the ability of motor-induced detachment to remove motors from very crowded MTs. 
	Here, the flow of motors is suboptimal due to the emergence of traffic jams, as in the case of vehicular traffic~\cite{Nagel1992}. 
	A decrease in the motor density may therefore enhance the numbers of motors transported along the MT per unit time. 
	We will see later that the existence of a maximum motor current follows naturally from the non-monotonic current-density relation, Eq.~(\ref{eq_j}).
	As an aside, one may thus speculate that motor-induced detachment may serve to optimise cargo transport along MTs by reducing crowding. 

	In this work, we are mainly interested in examining the collective dynamics of kinesin-1~\cite{Vale1985}. 
	In experiments, such as those in the study of Telley et al.~\cite{Telley2009}, its collective motion has been characterised in terms of run length on the MT $l$, dwell time $\tau$, velocity $V$, and the rate $\lambda$ (the number of motor landings on the MT per unit time and length). 
	All of these quantities may also be extracted from simulation data.
	However, not all of the model parameters necessary for simulations can be directly measured in experiments. 
	We will therefore employ the following strategy: First, we develop a theoretical analysis of our model, and extract model parameters from experimental data as far as possible.
	With analytical expressions for all relevant quantities at hand, we then fit our model to the experimental measurements. 
	Eventually, we will show that, with the global fit parameters obtained in this way, the theoretical predictions and simulation data of our model are in excellent agreement with experimental measurements.

\subsection*{Mean-field theory} 

	The configuration of a lattice at any given instant in time is described by a set of occupation numbers $\{ n_i \}$. 
	A lattice site $i$ (a tubulin heterodimer on the protofilament) is either empty ($n_i \,{=}\, 0$) or occupied by the front head ($n_i \,{=}\, f$) or back head ($n_i \,{=}\, b$) of a motor dimer. 
	For a statistical description we need the one-site and two-site probabilities, defined as 
\begin{subequations}
\begin{align}
	p(i,\alpha) &= \Prob (n_i \,{=}\, \alpha) ~,\\
	p(i,\alpha;j,\beta) &= \Prob (n_i \,{=}\, \alpha \wedge n_j \,{=}\, \beta) ~.
\end{align}
\end{subequations}
	We denote the position of a motor by the position of its front head and define the time-averaged dimer density as 
\begin{equation}
	\rho_i = p(i,f) ~,
\end{equation}
	which is then bounded to $\rho \in [0,\frac12]$.

	The rate of change of these probabilities can be described in terms of a set of master equations~\cite{Weber2016}. 
	For instance, for the time evolution of the probability that site $i$ is occupied by the front head of a motor, one obtains
\begin{align}
	\partial_t & p(i,f)= \nu \, \bigl[ p(i{-}1,f;i,0) - p(i,f;i{+}1,0) \bigr] \label{eq_dtp} \\
	 & + \omega_A \, p(i,0;i{-}1,0) - \omega_D \, p(i,f) - \theta \, p(i,f;i{+}1,b)~. \nonumber 
\end{align}
	Here, the first term on the right-hand side represents a transport current given by the difference between a gain and a loss term. 
	The gain term describes the probability per unit time that a motor (front head of a dimer) located at lattice site $i{-}1$ moves forward onto an empty site $i$, and the loss term describes the probability per unit time that a motor hops from site $i$ to the next (empty) site, $i{+}1$. 
	The remaining terms describe attachment and detachment processes with the joint probabilities selecting the allowed lattice configurations. 
	Thus, attachment of a dimer to the lattice is possible only if two neighbouring empty sites are available ($n_i \,{=}\, 0$ and $n_{i-1} \,{=}\, 0$). 
	While an interaction-induced detachment process requires that two dimers are immediate neighbours ($n_i \,{=}\, f$ and $n_{i+1} \,{=}\, b$), the rate of spontaneous detachment is proportional to the single-site probability $p(i,f)$.

	In general, the master equation, Eq.~(\ref{eq_dtp}), is not closed as it links single-site to two-site joint probabilities. 
	However, progress can be made by employing a mean-field approximation that neglects all correlations between the positions of motor dimers other than the steric constraint that dimers are not allowed to overlap, i.e. the front and the back heads of different motors cannot occupy the same lattice site. 
	Furthermore, for rigid dimers $n_i \,{=}\, b$ implies that site $i{+}1$ is occupied by the front head of the same motor, $n_{i{+}1} \,{=}\, f$.

	In order to show how the two-site joint probabilities can be reduced to one-site probabilities we will consider as an example $p(i,f;i{+}1,b)$.
	This probability, like any joint probability, can be expressed in terms of a conditional probability: $p(i,f;i{+}1,b) \,{=}\, p(i{+}1,b \vert i,f) \, p(i,f)$.
	As we are neglecting correlations in the position of different dimers, the probability that site $i{+}1$ is occupied by the back head of a dimer is independent of whether site $i$ is occupied by the front head of another dimer or empty: $p(i{+}1,b \vert i,f) $ $= p(i{+}1,b \vert i,0)$. 
	Hence, in a mean-field approximation we have $p(i{+}1,b \vert i,f) $ $= p \left( i{+}1,b \vert (i,f) {\vee} (i,0) \right) $ $= p \left( i{+}1,b \vert {\neg} (i,b) \right) $. 
	Using Bayes' theorem, this can be rewritten in the form $p \left( {\neg} (i,b) \vert  i{+}1,b \right) $ $\times\, p(i{+}1,b) / p \left( {\neg} (i,b) \right)$. 
	Here, the remaining conditional probability $p \left( {\neg} (i,b) \vert  i{+}1,b \right) $ equals $1$ because the states $(i, b)$ and $(i{+}1,b)$ are mutually exclusive. 
	Hence, we are left with the desired decomposition into single-site occupation probabilities:
\begin{equation}
	p(i,f; i{+}1,b) \,{=}\, \frac{p(i{+}1,b) \, p(i,f)}{1{-}p(i,b)} {=}\frac{ p(i{+}1,b) \, p(i,f) }{p(\neg( i,b))} ~.
\end{equation}
	Compared to a naive decomposition into single-site occupation probabilities $p(i{+}1,b) p(i,f)$, this equation includes a factor $1{-}p(i,b)$ which corrects for dimers spanning sites $i$ and $i+1$, i.e., which takes into account those correlations that are due to the dimeric nature of the motor molecules. 
	In the following we refer to such a factor as the local correlation factor. Using  $p(i,b) \,{=}\, p(i{+}1,f)$ one may rewrite this result solely in terms of the density $\rho_i$ as  
\begin{equation}
	p(i,f; i{+}1,b) \,{=}\, \frac{\rho_{i{+}2} \, \rho_i}{1{-}\rho_{i{+}1}} ~.
\end{equation}

	In the same way, cf. Ref.~\cite{Pierobon2006a}, we can also approximate the other joint probabilities of Eq.~(\ref{eq_dtp}). The ensuing mean-field master equation reads
\begin{align}
	\partial_t & \rho_i = \nu \left[ \frac{(1{-}\rho_i{-}\rho_{i{+}1}) \rho_{i-1}}{1{-}\rho_i} - \frac{(1{-}\rho_{i{+}1}{-}\rho_{i{+}2})\rho_i}{1{-}\rho_{i{+}1}} \right] \label{eq_dtrhoi} \\
	& + \omega_A \frac{( 1{-}\rho_i{-}\rho_{i{+}1} )( 1{-}\rho_{i{-}1}{-}\rho_i )}{1{-}\rho_i} 
	  - \omega_D \rho_i - \theta \frac{\rho_{i{+}2} \, \rho_i}{1{-}\rho_{i{+}1}} ~.
	\nonumber 
\end{align}
	In the stationary state, where $\partial_t \rho_i \,{=}\, 0$, this expression recursively determines the occupation density of site $i$ in terms of the densities of the neighbouring sites $i {\pm} 1$.
	In general, the dynamics of such a system is very rich and entails boundary-induced phase transitions~\cite{Lakatos2003a,Shaw2003,Parmeggiani2004,Pierobon2006a,Vuijk2015a}. 

	As discussed above, kinesin-1 has a relatively short run length and our focus is the bulk of MTs. Hence, we may assume that the motor density is constant, $\rho_i \,{=}\, \rho$, and arrive at the mean-field equation
\begin{equation}
	\partial_t \rho
	= \omega_A \frac{(1 {-} 2 \rho)^2}{1 {-} \rho} - \omega_D \rho - \theta \frac{\rho^2}{1{-}\rho}~,
	\label{eq_dtrho}
\end{equation}
	which yields the motor density $\rho_s$ in the stationary state ($\partial_t \rho=0$) as
\begin{equation}
	\rho_s =\frac{2 \omega _A}{4 \omega _A{+}\omega _D{+}\sqrt{4 \omega _A \omega _D{+}4 \theta  \omega _A{+}\omega _D^2}}~. 
	\label{eq_rhodimer}
\end{equation}
	Note that we could also have arrived at Eq.~(\ref{eq_rhodimer}) by assuming attachment-detachment balance
\begin{equation}
	\omega_A \, p(i,0;i{-}1,0) = \omega_D \, p(i,f) + \theta \, p(i,f;i{+}1,b)~.
\end{equation}
	As we are only interested in the behaviour at steady state, we will omit the index $s$ in the following, i.e. $\rho := \rho_s$.

	By employing the mean-field approximation we can also derive an expression for the motor current $j$. This quantity is defined as the number of motors that pass through a site on the MT per unit time, and is therefore given by $j_i \,{=}\, \nu p(i,f;i{+}1,0)$.
	By analogy with the derivations of the previous paragraph and Ref.~\cite{Pierobon2006a}, the motor current simplifies to 
\begin{equation}
j(\rho) \approx \nu \, \frac{\rho \, (1 {-} 2 \rho)}{1 {-} \rho} ~. \label{eq_j}
\end{equation}
	In this equation, we again identify the local correlation factor $1/(1{-}\rho)$. 
	Its significance can be understood as follows: Compared to the current-density relation for monomeric particles, $j(\rho) \,{=}\, \rho(1{-}\rho)$, Eq.~(\ref{eq_j}) is skewed, i.e. its maximum lies at a density exceeding half-occupation, $\rho \,{=}\, \frac12 (2{-} \sqrt{2}) \,{\approx}\, 0.29$. 
	This agrees remarkably well with the intuitive value for the density $\frac13$, where on average, every dimer is followed by a vacancy, and is therefore free to jump. 

	With the analytical expressions for the stationary motor density $\rho$ on the MT [Eq.~(\ref{eq_rhodimer})] and their flux $j(\rho)$ [Eq.~(\ref{eq_j})], we now have a description of the most central physical quantities that characterise the collective motion of molecular motors on a MT. 
	As Figs.~\ref{sf_general_rho} and \ref{sf_general_j} show, these analytically calculated quantities agree very well with data from stochastic simulations.

	Unfortunately, with present-day experimental techniques, it is difficult to measure collective quantities like the density $\rho$ and the current $j$.
	It is much easier to determine quantities derived from the observation of single labelled motors. 
	These include the dwell time $\tau$ of motors on the MT, their velocity $V$, run length $l$, and the landing rate $\lambda$. 
	In order to define the link between theory and experiment which we ultimately aim for, we must therefore also find expressions for these quantities.

	We first turn to the calculation of the dwell time $\tau$. 
	A motor located at site $i$ can detach either spontaneously at rate $\omega_D$, or additionally at a rate $\theta$ when another motor is located right next to it at site $i{+}2$. 
	The corresponding probability is given by $p(i{+}2,f \vert i,f)$, which reduces to $\rho/(1{-}\rho)$, following the same steps as before. 
	Hence, the dwell time is given by the inverse of the total detachment rate, comprising spontaneous and interaction-induced detachment:
\begin{equation}
	\tau \approx \left[ \omega_D + \theta \frac{\rho}{1{-}\rho} \right]^{-1} ~.
	\label{eq_tau}
\end{equation}
	Similarly, in order to obtain the velocity of a motor we need to consider the probability that a particle located at site $i$ finds the next site empty, $p(i{+}1,0 \vert i,f)$. 
	This gives for the motor velocity, again using a mean-field approximation,
\begin{equation}
	V
	= V_0 \, p(i{+}1,0 \vert i,f) 
	\approx V_0 \, \frac{1{-}2 \rho}{1{-}\rho} ~.
	\label{eq_v}
\end{equation}
With Eqs.~(\ref{eq_tau}) and (\ref{eq_v}), the run length of a motor is given by 
\begin{equation}
	l = \tau V \approx V_0 \, \frac{1{-}2\rho}{\omega_D (1{-}\rho) + \theta \rho} ~.
\end{equation}

	Finally, we need to compute the landing rate of kinesin on a MT. 
	In experiments, this quantity is determined by labelling only a small fraction of kinesin, e.g., with GFP, while the vast majority of motors remains unlabelled~\cite{Telley2009}. 
	The concentration of labelled motors is kept constant at a reference concentration $c_0$, and the unlabelled motors act as crowding agents which are added at varying concentrations.
	The landing rate is then obtained by counting how many labelled motors land on the MT per unit length and time.
	In our model, a motor can attach to a site $i$ on the MT only if it finds both site $i$ and the adjacent lattice site $i{-}1$ empty, $n_i \,{=}\, n_{i{+}1} \,{=}\, 0$. 
	With $\lambda_0$ being the landing rate of the normalised amount ($c_0$) of labelled kinesin on an otherwise empty MT, the landing rate is $\lambda \,{=}\, \lambda_0 \, p(i,0;i{-}1,0)$, which at the mean-field level is approximated by 
\begin{equation}
	\lambda \approx \lambda_0 \frac{(1-2 \rho)^2}{1-\rho} ~.
	\label{eq_landing}
\end{equation}
	It is important to note that the normalised landing rate $\lambda_0$ may differ from $\omega_A(c_0)$.
	This is because the size of a label such as GFP is comparable to that of the motor. Hence, the attachment rates of labelled and unlabelled motors to the MT may be different.

\begin{figure*}[hbt]
	\includegraphics[width=.99\textwidth]{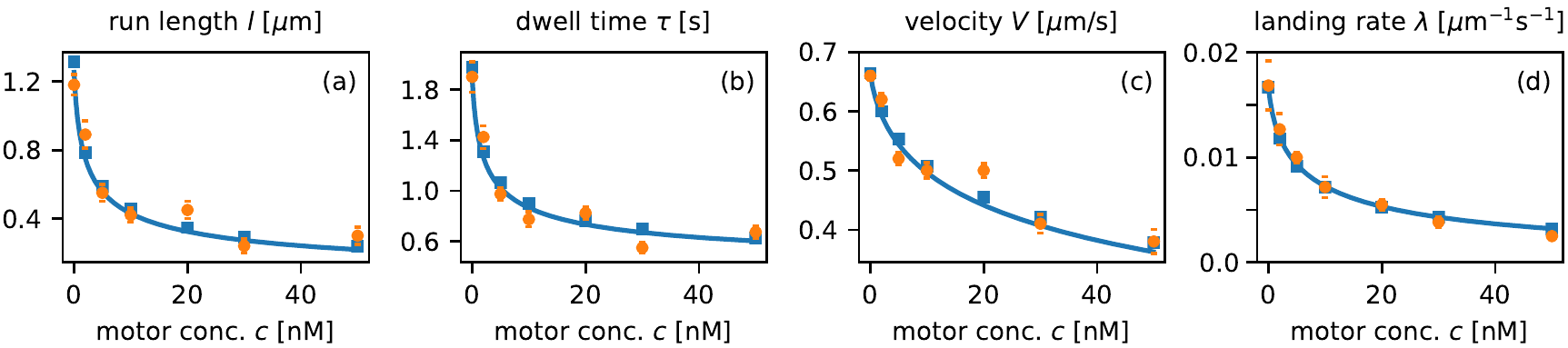}
	\subfloat{\label{sf_l}}
	\subfloat{\label{sf_tau}}
	\subfloat{\label{sf_v}}
	\subfloat{\label{sf_lambda}}
	\caption{
		\textbf{Comparison with experimental data.}
		Orange circles show the measurements for \protect\subref{sf_l} the run length, \protect\subref{sf_tau} dwell time, \protect\subref{sf_v} velocity, and \protect\subref{sf_lambda} landing rate of kinesin motors, as measured by Telley et al.~\cite{Telley2009}.
		In blue, we show the fit of our model to this data. Lines are results of our mean field theory, squares compare these calculations with simulations based on Gillespie's algorithm.
		\label{fig_fit_dimer}}
\end{figure*}
\begin{figure*}[hbt]
	\includegraphics[width=.99\textwidth]{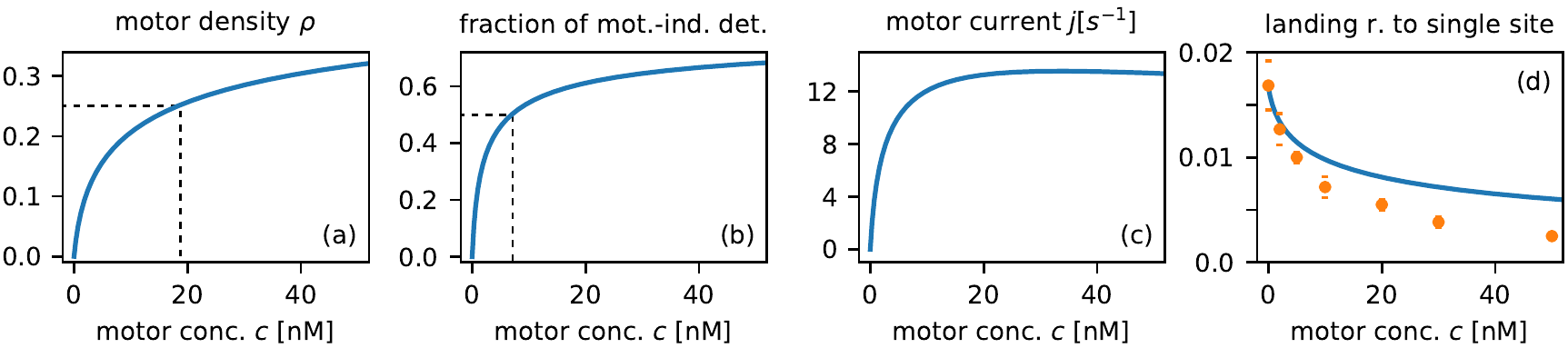}
	\subfloat{\label{sf_density}}
	\subfloat{\label{sf_rel_spont_det}}
	\subfloat{\label{sf_current}}
	\subfloat{\label{sf_lambda_s}}	
	\caption{
		\textbf{Characterisation of crowding effects.}
		The plot depicts important physical quantities available from our model, for the same parameters as in Fig.~\ref{fig_fit_dimer}.
		\protect\subref{sf_density} The density of motors on the MT. Because kinesin-1 is a dimer, $\rho \,{=}\, \frac12$ implies that the lattice is fully decorated with motors.
		\protect\subref{sf_rel_spont_det} Fraction of detachment events which are due specifically to motor-induced detachment. 
		Even at low concentrations around $7~\nM$, facilitated dissociation is as prominent as spontaneous detachment.		
		\protect\subref{sf_current} The motor current on the MT, i.e., the number of motors passing over a lattice site per unit time.
		\protect\subref{sf_lambda_s} The landing rate of motors on the MT (orange: experimental data~\cite{Telley2009}, blue: mean-field results), assuming that a single lattice site were sufficient for the landing of a motor. The agreement is worse than for the original model [Fig.~\ref{sf_lambda}].
		\label{fig_model_predictions}
		}
\end{figure*}

\subsection*{Comparison with experimental data}
	The primary goal of this work is to compare the predictions of our theoretical model with experimental data. 
	Telley et al.~\cite{Telley2009} have provided an extensive set of measurements for the motor kinesin-1, which is shown in Fig.~\ref{fig_fit_dimer}. 
	Here, the volume concentration of the motor is varied, and this process is incorporated into our model by setting $\omega_A \,{=}\, \omega_a c$.
	From their data, we can directly extract several of our model parameters.
	The hopping rate $\nu$ is obtained from the velocity $V_0$ of a motor in the limit of low motor density (Fig.~\ref{sf_v}),
\begin{equation}
	\nu = 0.66 \, \mum \, \sec^{-1} a^{-1}=79 \, \sec^{-1} ~. 
	\label{eq_nu} 
\end{equation}
	The detachment rate $\omega_D$ follows from the dwell time at small motor concentration (Fig.~\ref{sf_tau}),
\begin{equation}
	\omega_D = \frac{1}{1.9 \, \sec}
	         = 0.53 \, \sec^{-1}~, 
\label{eq_omegad}
\end{equation}
	and similarly the landing rate of a normalised amount of labelled kinesin can be directly read off from Fig.~\ref{sf_lambda} at $c \,{\approx}\, 0$,
\begin{equation}
	\lambda_0 = 1.8 \cdot 10^{-2} \mum^{-1} \sec^{-1} \label{eq_lambda0}~.
\end{equation}

	This leaves two parameters to be specified, the attachment rate of unlabelled motors to the MT per concentration, $\omega_a$, and the rate $\theta$ specifying interaction-induced detachment. 
	As there are four independent sets of quantities that have been measured~\cite{Telley2009} (run length, dwell time, velocity, and landing rate), comparison of all four with our theoretical results constitutes a stringent test of the validity of the assumptions on which the model is based.
	We have performed a global fit for the four independent quantities $l$, $\tau$, $V$, and $\lambda$ by minimising the squared sum of deviations between experimental measurements and mean-field results, weighted by the experimental confidence interval, see Methods. 
	This gives the following values for the rates
\begin{subequations}
	\begin{align}
	\omega_a & =5.4 \cdot 10^{-2} \, \textnormal{nM}^{-1} \sec^{-1}~, \label{eq_omegaa} \\
	\theta & =2.4 \,\sec^{-1} ~. \label{eq_theta}
	\end{align}
	\label{eq_fitparams}
\end{subequations}
	As can be seen in Fig.~\ref{fig_fit_dimer}, using these global fit parameters we find excellent agreement between our theory and all experimentally measured quantities. 

	Both these fit parameters are interesting in themselves. 
	The attachment rate $\omega_a$ specifies how quickly kinesin attaches to empty lattice sites. 
	In this context, one must keep in mind the fact that the physical quantity underlying the fit is the total motor density $\rho$ on the MT, while the data from Telley et al.~\cite{Telley2009} are derived from observations of the small minority of labelled motors.
	In our model, the rate $\omega_a$ specifies the attachment rate of the unlabelled motors, which act as a crowding agent but are otherwise invisible experimentally~\cite{Telley2009}. 
	How then does $\omega_a$ compare to the landing rate $\lambda_0$ for labelled motors? 
	This rate was measured at a motor concentration of $5$ pM and, assuming that motors in the TIRF setup can walk on roughly half of the $13$ protofilaments~\cite{Schneider2015}, this can be converted into a per-site attachment rate of approximately $5 \cdot 10^{-3} \, \nM^{-1} \sec^{-1}$. 
	This value is 10 times smaller than the attachment rate for unlabelled motors, and it demonstrates that, while labelling with GFP conserves many kinetic parameters of native kinesin~\cite{Block2003a,Seitz2006a,Guydosh2009}, the attachment rate of the labelled protein is significantly lower.
	
	Secondly, let us look more closely at the rate $\theta$, which quantifies motor-induced detachment from the filament. 
	The value of $\theta$ exceeds that of the spontaneous detachment rate $\omega_D$ by four-fold. 
	This is remarkable, because it implies that, under crowded conditions, motor-induced detachment is the dominant mechanism by which motors leave the MT. 
	We will analyse this and other implications of these parameters in greater detail in the following section.

\subsection*{Analysis of crowding effects}

	One strength of our approach to the quantitative description of the collective dynamics of molecular motors with a theoretical model is that it allows us to infer physical quantities which are experimentally difficult to access. 
	In particular, it is interesting and instructive to study the behaviour of the motor density along the MT, $\rho$, which is the fundamental quantity characterising the degree of crowding on the MT.
	In Figure~\ref{sf_density}, $\rho$ is plotted as a function of the volume concentration of motors $c$. 
	At small concentrations, the density rises steeply with $c$, and becomes half-maximal around $20~\nM$. 
	At this concentration, on average every second binding site on the MT is occupied by a motor head. 
	As $c$ is increased further, the motor density rises only modestly. 
	This is because attachment of additional motors becomes increasingly unlikely when many motors are already present on the MT, and motor-induced detachment becomes more prominent.
	
	Figure~\ref{sf_rel_spont_det} shows the fraction of motor detachments induced by the presence of another motor, plotted as a function of $c$. 
	With Eq.~(\ref{eq_dtrho}), we find that the contributions of spontaneous and motor-induced detachment are already comparable at a motor concentration around $7~\nM$, significantly below the concentration required for half-occupation [Fig.~\ref{sf_density}]. 
	The reason for this is that the rate $\theta$ exceeds $\omega_D$ by several-fold, such that motor-induced detachment plays the central role even on filaments with relatively little crowding.
	The steep increase in the contribution of motor-induced detachment to all dissociation events at low motor concentrations also explains the rapid decrease of quantities such as the motors' run length $l$ [Fig.~\ref{sf_l}] and dwell time $\tau$ [Fig.~\ref{sf_tau}] at these concentrations.

	The motor current $j$ may also be examined directly with our model and the parameters extracted from experimental measurements (Fig.~\ref{sf_current}).
	Once more, we find a steep increase at low concentrations.
	The current becomes maximal at around $c  \,{ \,{\sim}\, }\,  20$ nM, i.e. the concentration where the density is half-maximal, and for higher concentrations the motor current remains almost constant.

	Finally, the good agreement of our model with experimental data allows us to study the impact of model variations.
	For example, it has been suggested~\cite{Mickolajczyk2015} that kinesin-1 first binds via a single head to the MT on landing, and subsequently attaches its other head.
	We have directly tested how a different attachment mechanism might affect the landing rate by assuming that a single binding site is sufficient for the motor to attach to the MT.
	As a result, the attachment term in Eq.~(\ref{eq_dtrho}) reduces to $\omega_A(1-\rho)$.
	Fig.~\ref{sf_lambda_s} compares the landing rate obtained in this way with experimental data.
	Clearly, neither with the fit parameters for the original model, nor with parameters fitted to the modified model do we obtain satisfactory agreement between theoretical results and experimental data.
	Therefore, our data suggest that kinesin can land on the MT only where two adjacent binding sites are empty.

\subsection*{Crowding alone does not lead to periods of no or slow motion of motors}
\begin{figure*}[hbt]
	\centering
	\includegraphics[width=.99\textwidth]{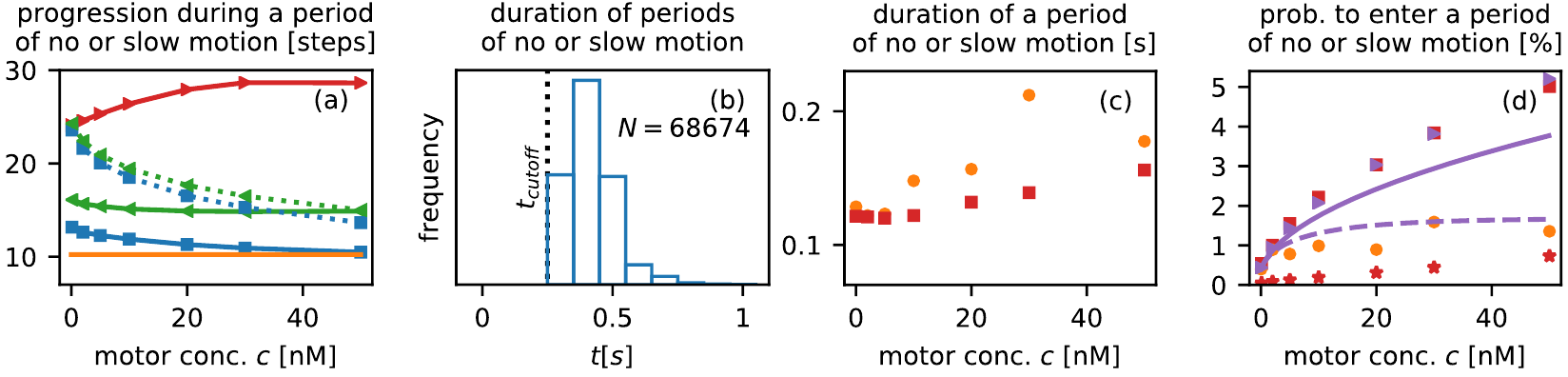}
	\subfloat{\label{sf_waitingprogression}}
	\subfloat{\label{sf_waitingtimedistribution}}
	\subfloat{\label{sf_pauseduration}}
	\subfloat{\label{sf_waitingprob_alternative}}
	\caption{
		\textbf{Periods in which kinesin-1 motors show no or only very slow motion.} 
		The model parameters are the same as in Fig.~\ref{fig_fit_dimer}. Experimental measurements from Telley et al.~\cite{Telley2009} are shown in orange.
		For the detection of these periods, the protocol of Telley et al.~\cite{Telley2009} was used (see the main text).
		\protect\subref{sf_waitingprogression}--\protect\subref{sf_waitingtimedistribution} Our model cannot explain the frequent periods of no or slow motion of motors observed experimentally. 
		\protect\subref{sf_waitingprogression} Distance travelled by a motor between the beginning and end of such a period. 
		A threshold value of $d_c=5$ sites (red line) is too large for reliable detection of these periods: Motors traverse for almost $30$ lattice sites between the beginning and end of such a period, which is three times the experimental result (orange line). 
		This implies that most of the detected events actually reflect stochastically slow motion which is otherwise normal, and hence the scheme detects these events inaccurately with this choice of $d_c$. 
		Reduction of the threshold to $d_c=3$ sites (green), or $d_c=2$ sites (blue) leads to results that are in closer agreement with experimental data. 
		However, this correspondence deteriorates on addition of Gaussian noise ($\sigma=20$nm) to the simulation data before applying the detection protocol (dashed lines)
		\protect\subref{sf_waitingtimedistribution} 
		Duration of the so detected periods of no or slow motion for $c=20$ nM and a detection threshold $d_c=2$ sites. 
		In contrast to the experimental findings, where an exponential distribution was observed, the duration peaks around $0.4$ s.
		This result does not change qualitatively when $d_c$ is varied, or Gaussian noise is added at various strengths. 
		\protect\subref{sf_pauseduration}--\protect\subref{sf_waitingprob_alternative} Qualitative agreement with experiments is found when motors can spontaneously switch between an active and inactive mode at rates extracted from experimental data~\cite{Telley2009}. 
		\protect\subref{sf_pauseduration} The duration of periods of no or slow motion detected from simulations of this model variant (red squares) is similar to those measured in experiments~\cite{Telley2009}.
		The duration was computed by extrapolating the (now) approximately exponential distribution of the detected periods below the cutoff time $0.3$ s~\cite{Telley2009} and the main text for details.
		\protect\subref{sf_waitingprob_alternative} The per-step probability that a motor is found in a period of no or slow motion. 
		The direct yield from the detection algorithm (red asterisks) is below experimentally observed~\cite{Telley2009} values. 
		When their frequency is corrected for the time cutoff (red squares), similar to the procedure used in experiments~\cite{Telley2009}, good qualitative agreement is found, in particular at low motor concentrations. 
		Violet triangles show the frequencies obtained with a different algorithm which counts motors that are inactive, or caught up in a traffic jam behind an inactive motor, but not motors which move slowly due to their stochastic motion. 
		The good agreement between these results and the original detection protocol (red squares) reveals that spontaneously inactivated motors are the dominant contribution for periods of no or slow motion.
		Solid and dashed lines show heuristic estimates of the probability of entering such a period, assuming that motors in a traffic jam require two, or only one binding site on the lattice. 
		For details, see the main text.
		\label{fig_pauses}
	}
\end{figure*}

	As shown in the previous sections, our mathematical model explains the kinetic data for the run length, dwell time, velocity, and landing rate of kinesin-1 motors on MTs with high accuracy.
	These quantities are averaged over a large number of motors and characterise their collective transport along MTs very well. 
	However, with our model, as well as in experiments, quantities other than averages are also accessible, such as the statistics of individual steps of motors. 
	Such quantities are instructive, as they afford insight into the stochastic motion of kinesin at a deeper level. 
	A particularly interesting finding made by Telley et al.~\cite{Telley2009} in this respect was that kinesin-1 motors, which normally move at a speeds as high as $79$ steps/s along the MT under uncrowded conditions, sometimes show periods in which they rest on the MT or their motion is at least considerably slowed down. 
	These periods lasted for several tenths of a second, during which a motor would typically proceed by dozens of steps.
	It was found that the frequency of these periods increased with the volume concentration of kinesin, and hence with the degree of crowding on the MT~\cite{Telley2009}.
		
	However, the authors of that study were only able to image the motors every $0.1$ s, such that the localisation accuracy of kinesin-1 was of the same order of magnitude as the typical distance traversed between two measurements.
	Furthermore, because kinesin's stepping mechanism includes chemical reactions as well as diffusive motion, this motor is a stochastic stepper.
	Consequently, Telley et al.~\cite{Telley2009} were faced with the problem of robustly distinguishing periods of no (or very slow) motion,\footnote{
		Note that Telley et al.~\cite{Telley2009} use the term \enquote{pause} for periods in which no or little motion was detected, and they further distinguish between \enquote{wait} and \enquote{stop} for such events in which kinesin continued its run subsequent to the pause, or detached from the MT. In this work, we distinguish between the \emph{phenomenon} observed in experiments, which we will call \enquote{periods of no or slow motion}, and the \emph{cause} of these periods, which we term \enquote{pause} in the following.
	} 
	in which motors are assumed to hardly move at all, from stochastically slow motion which simply reflects the stochasticity of kinesin's steps but is otherwise normal.
	
	To overcome these problems, Telley et al.~\cite{Telley2009} developed a detection scheme for the periods of no or slow motion as follows:
	The location of the motors was measured every $0.1$ s. 
	If a motor failed to advance a critical distance $d_c$ between two time frames, or its motion was directed off-axis or backwards (exceeding a critical angle $\alpha_c$), this displacement was considered as a \emph{candidate} for the onset of a period of no or slow motion.
	However, in order to mark the start of such a period, three successive small displacements were required.
	To account for the effect of experimental noise, single advances exceeding $d_c$ were allowed during a period of no or slow motion, so that the period was only considered as terminated when the displacement was greater than $d_c$ twice in a row.
	
	The key parameters which determine the sensitivity of the detection of periods of no or slow motion are $\alpha_c$ and (in particular) $d_c$.
	On the one hand, these quantities should be chosen to be so large that fluctuations due to experimental noise are unlikely to prematurely terminate such periods. 
	On the other hand, the critical distance must be kept so small that these periods can be robustly distinguished from normal motion which is slow because of the stochasticity of kinesin's steps.
	With the parameters $d_c=40$ nm and $\alpha_c=60^\circ$, Telley et al.~\cite{Telley2009} found that approximately every second kinesin motor showed a period of no or slow motion at some point while it progressed along the MT.
	During the periods of no or slow motion, motors proceeded on average $10$ lattice sites.
	This value seems large, but it is much less than the expected $\sim 30$ lattice sites which a motor would traverse under uncrowded condition during the minimal time necessary for detection of these periods (0.3s).
	
	In order to compare the predictions of our model with the experimental data of Telley et al.~\cite{Telley2009}, we adapted and applied their experimental detection scheme for periods of no or slow motion to our system.
	Note, however, that the motion of motors is restricted to a single dimension in our model, while occasional side-steps, as well as off-axis fluctuations are possible in experiments. 
	Consequently, the two parameters, $d_c$ and $\alpha_c$, used for the experimental detection have to be reduced to a single parameter $d_c$ for our purposes. 
	Moreover, since a finite progression $d_c$ between two frames was allowed primarily in order to account for experimental inaccuracies which are absent in simulations, $d_c$ has to be critically evaluated, and the role of noise must be simulated.
	To this end, we first chose the same threshold distance $d_c=40$ nm as in Ref.~\cite{Telley2009}, corresponding to $5$ lattice sites.
	With this value, we found that the progression of a motor between the beginning and the end of a so defined period of no or slow motion was almost $30$ lattice sites [Fig.~\ref{sf_waitingprogression}]. 
	This is significantly larger than the experimentally measured length of $10$ lattice sites, and therefore indicates that most of the detected events in fact do not show behaviour which is physically different from normal motion. 
	Thus, most of the periods of no or slow motion detected with this choice of $d_c$ result from the stochastic motion of kinesin.
	Even for a threshold distance of $3$ lattice sites, the progression exceeded experimental data, so that we had to reduce the value of $d_c$ to $2$ lattice sites in order to find agreement with experimental results [Fig.~\ref{sf_waitingprogression}].
	However, the agreement found with this parameter choice deteriorated when Gaussian noise was added to the simulation data (in order to account for experimental fluctuations) before applying the protocol [$\sigma=20$ nm in Fig.~\ref{sf_waitingprogression}].

	Moreover, the statistics of the durations of periods of no or slow motion detected from our simulation data differed from experimental results. 
	While Telley and coworkers~\cite{Telley2009} report an exponential distribution, our results indicate a non-exponential distribution with peaks around $0.4$--$0.5$ s, see Fig.~\ref{sf_waitingtimedistribution}. 
	Also the addition of Gaussian noise, or variation of the detection threshold $d_c$ did not qualitatively change this distribution.	
	
	We therefore conclude that the detection protocol of Telley et al.~\cite{Telley2009} is inappropriate for the analysis of the data obtained from stochastic simulations of our original model for two reasons.
	Firstly, it fails to distinguish reliably between periods of no or slow motion and stochastically slow, but normal motion of kinesin, as the progression of motors between the beginning and end of the detected periods clearly exceeds experimental results.
	Secondly, the distribution of the durations of periods of no or slow motion in simulations differs fundamentally from the experimental findings of Telley et al.~\cite{Telley2009}.
	Consequently, in order to understand the full dynamic behaviour of motors on the MT additional stochastic processes must be taken into account, which are not captured by our original model.
	This will be the focus of the next section.
	
	\subsection*{Spontaneous pausing of motors leads to crowding-dependent frequencies of periods of no or slow motion}
	In order to examine model variations which could possibly explain the experimental findings of Telley et al.~\cite{Telley2009} on periods in which the motors did not or only very slowly move, we looked at the data they obtained at low motor concentrations.
	Interestingly, even though motors proceed along the MT (almost) in the absence of other motors at these concentrations, periods of no or slow motion were observed occasionally. 
	This prompted us to study a variant of our model in which motors can stochastically pause on the MT, i.e., they may temporarily switch to an inactive state in which they cannot move.
	From the experimental data at low concentrations, we read off a per-step chance of lapsing into inactivity of $p_\textnormal{inactivation}=0.4\%$, and a pausing time with average duration $T = 0.12$ s, after which motors are reactivated again. 
	We therefore introduced rates $r_\textnormal{inactivation}=0.004 \nu=0.32 \, \sec^{-1}$ and $r_\textnormal{activation}=1/0.12 \, \sec^{-1}=8.3 \,\sec^{-1}$ at which motors switch to an inactive or active state, respectively. 
	At the molecular level, a motor might become inactive, for instance, when a motor is trapped in an unfavourable chemical state due to imperfect synchronisation of its heads~\cite{Mickolajczyk2015}; however, the particular molecular mechanism involved is not important for the argument below.

	If motors are allowed to switch into an inactive mode, we expect crowding to enhance the measured probability of undergoing a period of no or slow motion, because other motors will tend to form a traffic jam behind inactive motors.
	Although the motors caught up in the traffic jam are not intrinsically inactive, they are unable to progress until the inactive motor has become active again. 
	Therefore, crowding should amplify the impact of stochastic pausing and consequently lead to frequent periods in which kinesin motors show no or only slow motion along the MT.
	
	We tested these expectations directly by performing Monte Carlo simulations of this variant model. 
	Since the two additional stochastic processes, namely spontaneous inactivation and activation of motors, are rare events, we found that they have only a small impact on motor run lengths, dwell time, velocity and landing rate (data not shown). 
	In contrast, motor behaviour changed considerably at the level of individual steps: Unlike the case in our original model, Fig.~\ref{fig_model}, the durations of periods of no or slow motion were (approximately) exponentially distributed in the variant model, in accordance with experimental findings~\cite{Telley2009}\footnote{
		As reactivation from an inactive state is a one-step process, the distribution of the duration of periods of no or slow motion should be \emph{exactly} exponentially distributed in the absence of crowding and noise; this agrees with simulation data analysed with the detection algorithm of Telley et al.~\cite{Telley2009}. 
		As the degree of crowding increased due to additional motors on the MT, and as noise was added to the simulation data, the distribution gradually changed and was non-exponential for high crowding and noise level, albeit with an exponential tail for durations $>0.5$ s. 
		In order to comply with the procedure of Telley et al.~\cite{Telley2009}, we used the distribution's tail to fit an exponential function to the simulation data, as we extrapolated the distribution below the cutoff value $0.3$ s in order to obtain, e.g., the mean duration.
		}. 
	Following Telley et al.~\cite{Telley2009}, it is essential to extrapolate this exponential distribution below the cutoff time $0.3$ s in order to obtain the corrected frequency and mean duration of the periods of no or slow motion.
	The reason for this is that the cutoff $0.3$ s is a technical choice, but there is no physical reason why motors would not also experience periods of no or slow motion which are shorter than that.
	As a result, periods of no or slow motion comprise the detected periods (those lasting $0.3$ s and longer), as well as the undetected periods (those of shorter duration).
	The mean duration of the periods of no or slow motion is therefore given by the parameter of the exponential decay of the distribution.
	Figure \ref{sf_pauseduration} shows the concentration dependence of the mean duration of periods of no or slow motion, as they were extracted from simulation data in this way, and they reproduce the experimental findings~\cite{Telley2009} well.
	Moreover, these values were almost independent of the parameter $d_c$ used for the detection algorithm, which ensures that periods of no or slow motion of our model variant are now detected robustly and accurately.
	
	We are now in a position to compare simulation data for the frequencies of periods of no or slow motion with those of the experiments of Telley et al.~\cite{Telley2009}, as shown in Fig.~\ref{sf_waitingprob_alternative}.
	While the uncorrected probabilities (asterisks) remain below experimental values, as expected, the frequencies corrected for the cutoff (squares) are comparable to those found experimentally~\cite{Telley2009} for low concentrations. 
	However, as the concentration is increased, we found that the frequencies measured in our simulations exceed experimental values. 
	This points to the need for further modifications of our model.

	In principle, any additional interactions can be included into our model and data obtained from stochastic simulations. 
	However, a more instructive approach for our purposes is, however, to analyse the physical principles leading to periods of no or slow motion, and explore how exactly the inactivation of a single motor results in the formation of traffic jams which amplify the effect of pausing.
	To study this, we employed a different algorithm to detect periods of no or slow motion: Here, we only counted motors that were (i) inactive themselves, or (ii) trapped in a traffic jam behind an inactive motor. 
	In contrast, events in which motors moved slowly because they were caught up in a stochastically assembled traffic jam (in which no motor is inactive) were not taken into account.
	The frequencies of periods of no or slow motion obtained with this alternative algorithm (triangles in Fig.~\ref{sf_waitingprob_alternative}) agree well with those calculated with the original algorithm (squares in the same Figure).
	This implies that although stochastically arising traffic jams (in which no motor is inactive) slow down the collective motion of motors~\cite{Parmeggiani2004,Leduc2012}, they do not increase the incidence of periods of no or slow motion. 
	In contrast, these periods are predominantly due to the spontaneous (and transient) inactivation of motors and the associated formation of traffic jams behind these motors.
	
	Given that the dominant cause of periods of no or slow motion is the formation of traffic jams behind inactive motors, further insight can be gained by estimating theoretically the length of these traffic jams.
	Imagine that a motor pauses at some lattice site. 
	Then, the $n$-th motor behind this inactive motor is on average $n/\rho$ sites away from it. 
	Since each motor requires two binding sites on the MT, the $n$-th motor therefore typically has to travel $n/\rho-2n$ sites to reach the end of the traffic jam. 
	Hence, the time needed for the $n$-th motor to reach the end of the traffic jam may be estimated as $t(n)=(n/\rho{-}2n)/V$. 
	As a consequence, during the time $T$ required for reactivation of an inactive motor, a traffic jam containing $N_1=n (T) \,{=}\, T V/(\rho^{-1}{-}2)$ motors will form. 
	After the inactive motor has resumed its run, all the motors stuck in the traffic jam can start moving again one after another, so that it will typically take a time $N_1 \nu^{-1}$ before the original traffic jam has completely dissolved. 
	During this time, another $N_2=n(N_1 \nu^{-1})$ motors will have reached the end of the traffic jam, and more time will be needed until this additional traffic jam is dispersed, and so on. 
	Taking the sum over the number of motors caught in traffic jams found in this way, the number of motors $N=N_1+N_2+\dots$ which are ultimately affected by a single spontaneously pausing motor is consequently obtained from a geometric series, yielding
\begin{equation}
	N=\frac{Tv}{\rho^{-1}-2-V \nu^{-1}}~.
	\label{eq_pause_amplification}
\end{equation}
	This equation suggests that the effect of spontaneous pausing is considerably amplified by crowding. 
	While the \emph{cause} of traffic jams is the inactivation of a single motor, the \emph{phenomenon} detected with the scheme of Telley et al.~\cite{Telley2009} is also visible for $N$ other motors that are effectively caught in a traffic jam; consequently, $p_\textnormal{per. no/slow mot.} \,{=}\, p_\textnormal{inactivation} (1{+}N)$. 
	Figure~\ref{sf_waitingprob_alternative} shows the probability per step obtained in this way. 
	Given the level of the heuristic arguments, the agreement with simulation data is satisfactory.

	Having a theoretical estimate for the density dependence, and with Eq.~(\ref{eq_rhodimer}) also the concentration dependence, of the frequencies of periods of no or slow motion at hand, further model variations can now be tested in a relatively simple way. 
	For example, it seems plausible that motors align in a traffic jam very compactly, such that each motor requires a single lattice site on the MT only. 
	This would be in accordance with studies in which the decoration of MT sheets with immobilised dimeric kinesin was investigated, and it was found that kinesin binds to the MT via a single head only under certain conditions~\cite{Vilfan2001a,Moyer1998}. 
	For this model, the $n$-th motor behind an inactive motor would then have to travel further compared to the original (i.e., spaced) jamming model, namely $n/\rho-n$ sites. 
	In consequence, the term $\rho^{-1}{-}2$ in Eq.~(\ref{eq_pause_amplification}) would be modified to $\rho^{-1}{-}1$, and the amplification of spontaneous pausing changes accordingly.
	As shown in dashed lines in Fig.~\ref{sf_waitingprob_alternative}, the resulting per step probability of entering a period of no or slow motion reproduces the experimental concentration dependence~\cite{Telley2009} better than the original model in which motors align spaciously in a traffic jam.
	
	In conclusion, we have shown that spontaneous and transient inactivation of motors is the key to an understanding of the occurrence of periods of no or slow motion.
	The frequency of these periods is determined by the formation of traffic jams, in which motors (which are not intrinsically inactive themselves) cannot, or only slowly progress.
	However, we are at present unable to uniquely determine the precise mechanisms of jamming, and predict quantitatively how exactly they amplify the frequencies of periods of no or slow motion of molecular motors. 
	A central problem seems to be that periods of no or slow motion are relatively short-lived compared to the threshold time required to detect such an event.
	This implies that large numbers of these events remain undetected, and can only be resolved by extrapolating the duration distribution, as explained above. 
	As a consequence, we expect that the estimates of the frequencies of periods in which kinesin motors move only very slowly or come to a complete halt on the MT are subject to relatively large errors.
	It will in the future therefore be important to further investigate the origin of these periods; in particular, algorithms have to be developed which allow a more direct detection of short pauses, e.g., by increasing the frame rate of experiments. 
	Furthermore, direct visualisation of the inactive state would be highly informative.
	In summary, crowding is most probably not the underlying reason for periods of no or slow motion of motors, but acts as an amplifier to increase their frequency, although their ultimate cause is related to inactive states of kinesin motors.

\subsection*{The step cycle of kinesin has (at least) two slow transitions}

	Our findings concerning the motor-induced detachment of kinesin motors provide insight into their stepping cycle.
	We would like to emphasise first that none of the results presented in the previous sections depends on whether disengagement of the front or rear motor from the MT is enhanced by the presence of another motor. 
	Consequently, \enquote{bouncing off} (the rear motor detaches) and \enquote{kicking off} (the front motor detaches) interactions lead to identical results (data not shown). 
	In fact, there are experimental indications that it is the trailing motor which bounces off when it encounters another motor on the MT. 
	This was suggested by, among others, Telley et al.~\cite{Telley2009}, who used non-motile rigour mutants, in addition to wild-type kinesin-1. 
	Here, the tightly bound mutant motors act as obstacles on the MT, and the wild-type motors detach at an enhanced rate on encountering such an obstacle. 
	This would also suggest that when two wild-type kinesin motors come into contact on the MT, it is the trailing motor that is more likely to detach.

	At the molecular level, these indications enables us to associate the motor-induced unbinding process with a specific state in the mechanochemical cycle of kinesin. 
	This cycle comprises transitions between several states in which one or both kinesin heads are bound to the MT, and the two heads contain different bound nucleotides. 
	During the stepping cycle, kinesin passes through a state in which only a single head is bound to the MT. 
	This weakly bound state is reached after the back (i.e., the tethered) head is released from the MT, and the head that remains bound to the MT binds and hydrolyses ATP. 
	It is likely that this one-head-bound (1HB) state, in which the head attached to the MT is associated either with ADP or ADP$\cdot$P$_i$, is the state from which motors usually detach into the cytosol at finishing their run~\cite{Hancock1999,Milic2014}. 
	If the lifetime of this state is increased, kinesin should therefore also unbind at an enhanced probability.

	We hypothesise that the increase in the detachment rate seen when two motors occupy directly adjacent binding sites on the MT is directly related to this weakly bound state. 
	More specifically, when the rear motor's tethered head attempts to step to the next binding site, but finds this site occupied by another motor, the rear motor can leave its 1HB state only by stepping back (which is rare~\cite{Clancy2011}), or by waiting until the next site is vacated. 
	In this case, the back motor is \enquote{trapped} in a weakly bound state, and the detachment rate is enhanced accordingly. 
	We, therefore, interpret $\theta$ as the dissociation rate of kinesin from the 1HB ADP($\cdot$P$_i$) state. 
	This interpretation is also supported by measurements of the dissociation rate of single-headed kinesin motors which are artificially held in the ADP and ADP$\cdot$P$_i$ state, where rates of $3.7 \, \sec^{-1}$ and $3.8 \, \sec^{-1}$, were found, respectively~\cite{Hancock1999}; these measurements are remarkably similar to the value of $\theta$ obtained from Eq.~(\ref{eq_theta}).
	Following these arguments, the time fraction $f$ which a motor spends in the 1HB state during a normal step, may be determined from $\omega_D = f \theta$. 
	By direct comparison, we obtain $f = 0.22$, which implies that kinesin-1 remains in the 1HB ADP($\cdot$P$_i$) state for approximately 22$\%$ of the time needed to complete a stepping cycle.

	In summary, our findings suggest that the kinesin-1 step cycle comprises (at least) two transitions which are of similar duration, as opposed to a single rate-limiting step. 
	This is in agreement with a recent interpretation of the kinesin step cycle~\cite{Hancock2016}.
	We believe that our study will also help to reconcile conflicting results on the number and type of rate-limiting steps obtained from optical trapping experiments~\cite{Clancy2011,Andreasson2015}, dark-field~\cite{Isojima2016} and interferometric scattering~\cite{Mickolajczyk2015} microscopy experiments, as well as from measurements of the statistics of single motor runs~\cite{Verbrugge2009}.
	While the methods employed in most of these experiments give rise to much shorter length and time scales, labelling of the heads of motors, or applying force to them using an optical trap risks interfering with the step cycle. 
	The advantage of our analysis is that interference effects are minimised. 
	Therefore, crowding experiments~\cite{Telley2009} provide unique insight into a microscopic process by in a minimally invasive way.

\section*{Discussion and Conclusion}

	In this work, we have theoretically studied the impact of interactions between kinesin-1 motors on their motility and transport properties along microtubules.
	Based on experimental observations, we have generalised a lattice gas model~\cite{Parmeggiani2003,Parmeggiani2004} that has previously proven successful in explaining collective phenomena, such as the existence of traffic jams, which have recently been observed experimentally for kinesin-8~\cite{Leduc2012}, and kinesin-4~\cite{Subramanian2013}. 
	The generalised model includes the additional process of motor-induced detachment from the microtubule when one motor is directly adjacent to another, as well as the stochastic inactivation (pausing) of motors.
	With only two fit parameters, namely the rate of motor-induced detachment $\theta$, and the attachment rate of motors to empty lattice sites $\omega_A$, our model can account for four independent sets of measurements from \emph{in vitro} experiments~\cite{Telley2009} with kinesin-1 (Fig.~\ref{fig_fit_dimer}). 

	The level of agreement of our model with experimental data allows us to explore the origin of the relatively long periods during which motors hardly move along the MT at all, which have been observed in experiments~\cite{Telley2009}.
	We find that crowding alone cannot explain the high frequency of these periods ( Fig.~\ref{fig_pauses}).
	We therefore hypothesize that motors may stochastically switch into an inactive mode.
	Consequently, crowding leads to the formation of traffic jams behind inactive motors; these traffic jams significantly amplify the number of motors which pause on the filament, Eq.~(\ref{eq_pause_amplification}).
	Our findings suggest that motors might actually be aligned very densely in a traffic jam (Fig.~\ref{fig_pauses}) such that every motor occupies only a single tubulin dimer, in accordance with Ref.~\cite{Vilfan2001a}.
	By comparing the rates of motor-induced detachment and spontaneous unbinding, we find that kinesin-1 motors spend approximately $22 \%$ of their stepping cycle in a weakly bound state.
	Most probably, motor-induced detachment occurs when the rear motor is held in this state for a prolonged time when two motors are directly adjacent, and that its unbinding is therefore increasingly likely.

	Our approach to quantitatively model the dynamics of molecular motors enables us to investigate collective properties of kinesin-1 motors in a \enquote{real life} situation. 
	Firstly, in the experiments of Telley and coworkers~\cite{Telley2009}, on which our model is based, only a small fraction of motors was labelled.
	Secondly, insight into the interactions of motors with each other has been gained in our study without perturbing motor behaviour by applying forces etc.
	Our results enable us, for example, to compare the landing rates of labelled and unlabelled motors, and we have found that in fact labelled motors attach to the MT more slowly than unlabelled motors.
	This illustrates that the choice of a large label can have a crucial impact on certain quantities, and thus great care should be taken in interpreting experimental data.
	Most importantly, our model and the experiments of Telley et al.~\cite{Telley2009} provide unique insight into the stepping cycle of kinesin, which allows us to estimate the lifetime of a specific, weakly bound state. 
	The major drawback of our method is at once its greatest strength: Our approach is very indirect. 
	The application of forces to kinesin motors, e.g. by using optical traps~\cite{Milic2014,Andreasson2015}, as well as the attachment of large labels such as gold particles to kinesin heads~\cite{Mickolajczyk2015,Isojima2016} might have crucial influence on motor dynamics~\cite{Khataee2017}.
	Therefore, indirect methods~\cite{Verbrugge2009,Sozanski2015} such as the approach employed in this work are essential to confirm, and improve experimental results found by direct observation.

	Future studies, both theoretical and experimental, will have to examine more closely the formation and dissolution of traffic jams induced by the spontaneous inactivity of a motor, for example.
	In the same way, the spatial arrangement and conformation of motors in a traffic jam requires closer attention. 
	Such studies are essential to further improve our understanding of the role of interaction between molecular motors for the dynamics along cytoskeletal filaments. 
	This might have important implications for the biological function of such processes in the crowded environments within cells.

\section*{Acknowledgements}
{
	\small
	We thank Ivo Telley and Thomas Surrey for insightful discussion and sharing the experimental data of Telley et al.~\cite{Telley2009}.
	We also acknowledge support by the DFG via project B02 within SFB 863, as well as by the German Excellence Initiative via the program \enquote{NanoSystems Initiative Munich} (NIM).
	}

{
\small

\begin{thebibliography}{59}
\providecommand{\url}[1]{\texttt{#1}}
\providecommand{\urlprefix}{ }

\bibitem[Gupta et~al.(2006)Gupta, Carvalho, Roof, and Pellman]{Gupta2006}
Gupta, M.~L., P.~Carvalho, D.~M. Roof, and D.~Pellman, 2006.
\newblock {Plus end-specific depolymerase activity of Kip3, a kinesin-8
  protein, explains its role in positioning the yeast mitotic spindle}.
\newblock \emph{Nature Cell Biology} 8:913--923.

\bibitem[Varga et~al.(2006)Varga, Helenius, Tanaka, Hyman, Tanaka, and
  Howard]{Varga2006}
Varga, V., J.~Helenius, K.~Tanaka, A.~A. Hyman, T.~U. Tanaka, and J.~Howard,
  2006.
\newblock {Yeast kinesin-8 depolymerizes microtubules in a length-dependent
  manner}.
\newblock \emph{Nature Cell Biology} 8:957--962.

\bibitem[Varga et~al.(2009)Varga, Leduc, Bormuth, Diez, and Howard]{Varga2009}
Varga, V., C.~Leduc, V.~Bormuth, S.~Diez, and J.~Howard, 2009.
\newblock {Kinesin-8 Motors Act Cooperatively to Mediate Length-Dependent
  Microtubule Depolymerization}.
\newblock \emph{Cell} 138:1174--1183.

\bibitem[Melbinger et~al.(2012)Melbinger, Reese, and Frey]{Melbinger2012}
Melbinger, A., L.~Reese, and E.~Frey, 2012.
\newblock {Microtubule Length Regulation by Molecular Motors}.
\newblock \emph{Physical Review Letters} 108:258104.

\bibitem[Rank et~al.(2018)Rank, Mitra, Reese, Diez, and Frey]{Rank2018}
Rank, M., A.~Mitra, L.~Reese, S.~Diez, and E.~Frey, 2018.
\newblock {Limited Resources Induce Bistability in Microtubule Length
  Regulation}.
\newblock \emph{Physical Review Letters} 120:148101.

\bibitem[Goshima et~al.(2005)Goshima, Wollman, Stuurman, Scholey, and
  Vale]{Goshima2005}
Goshima, G., R.~Wollman, N.~Stuurman, J.~M. Scholey, and R.~D. Vale, 2005.
\newblock {Length Control of the Metaphase Spindle}.
\newblock \emph{Current Biology} 15:1979--1988.

\bibitem[Stumpff et~al.(2008)Stumpff, von Dassow, Wagenbach, Asbury, and
  Wordeman]{Stumpff2008}
Stumpff, J., G.~von Dassow, M.~Wagenbach, C.~Asbury, and L.~Wordeman, 2008.
\newblock {The Kinesin-8 Motor Kif18A Suppresses Kinetochore Movements to
  Control Mitotic Chromosome Alignment}.
\newblock \emph{Developmental Cell} 14:252--262.

\bibitem[Jordan and Wilson(2004)]{Jordan2004}
Jordan, M.~A., and L.~Wilson, 2004.
\newblock {Microtubules as a target for anticancer drugs}.
\newblock \emph{Nature Reviews Cancer} 4:253--265.

\bibitem[Vale et~al.(1985)Vale, Reese, and Sheetz]{Vale1985}
Vale, R.~D., T.~S. Reese, and M.~P. Sheetz, 1985.
\newblock {Identification of a novel force-generating protein, kinesin,
  involved in microtubule-based motility.}
\newblock \emph{Cell} 42:39--50.

\bibitem[Hirokawa and Takemura(2005)]{Hirokawa2005}
Hirokawa, N., and R.~Takemura, 2005.
\newblock {Molecular motors and mechanisms of directional transport in
  neurons}.
\newblock \emph{Nature Reviews Neuroscience} 6:201--214.

\bibitem[Hirokawa(1998)]{Hirokawa1998}
Hirokawa, N., 1998.
\newblock {Kinesin and Dynein Superfamily Proteins and the Mechanism of
  Organelle Transport}.
\newblock \emph{Science} 279:519--526.

\bibitem[Akhmanova and Steinmetz(2008)]{Akhmanova2008}
Akhmanova, A., and M.~O. Steinmetz, 2008.
\newblock {Tracking the ends: a dynamic protein network controls the fate of
  microtubule tips}.
\newblock \emph{Nature Reviews Molecular Cell Biology} 9:309--322.

\bibitem[Leduc et~al.(2012)Leduc, Padberg-Gehle, Varga, Helbing, Diez, and
  Howard]{Leduc2012}
Leduc, C., K.~Padberg-Gehle, V.~Varga, D.~Helbing, S.~Diez, and J.~Howard,
  2012.
\newblock {Molecular crowding creates traffic jams of kinesin motors on
  microtubules}.
\newblock \emph{Proceedings of the National Academy of Sciences}
  109:6100--6105.

\bibitem[Subramanian et~al.(2013)Subramanian, Ti, Tan, Darst, and
  Kapoor]{Subramanian2013}
Subramanian, R., S.-C. Ti, L.~Tan, S.~A. Darst, and T.~M. Kapoor, 2013.
\newblock {Marking and Measuring Single Microtubules by PRC1 and Kinesin-4}.
\newblock \emph{Cell} 154:377--390.

\bibitem[Vilfan et~al.(2001)Vilfan, Frey, Schwabl, Thorm{\"{a}}hlen, Song, and
  Mandelkow]{Vilfan2001a}
Vilfan, A., E.~Frey, F.~Schwabl, M.~Thorm{\"{a}}hlen, Y.-H. Song, and
  E.~Mandelkow, 2001.
\newblock {Dynamics and cooperativity of microtubule decoration by the motor
  protein kinesin11Edited by W. Baumeister}.
\newblock \emph{Journal of Molecular Biology} 312:1011--1026.

\bibitem[Muto et~al.(2005)Muto, Sakai, and Kaseda]{Muto2005}
Muto, E., H.~Sakai, and K.~Kaseda, 2005.
\newblock {Long-range cooperative binding of kinesin to a microtubule in the
  presence of ATP}.
\newblock \emph{The Journal of Cell Biology} 168:691--696.

\bibitem[{H Roos} et~al.(2008){H Roos}, Camp{\`{a}}s, Montel, Woehlke, Spatz,
  Bassereau, and Cappello]{Roos2008}
{H Roos}, W., O.~Camp{\`{a}}s, F.~Montel, G.~Woehlke, J.~P. Spatz,
  P.~Bassereau, and G.~Cappello, 2008.
\newblock {Dynamic kinesin-1 clustering on microtubules due to mutually
  attractive interactions}.
\newblock \emph{Physical Biology} 5:046004.

\bibitem[Telley et~al.(2009)Telley, Bieling, and Surrey]{Telley2009}
Telley, I.~A., P.~Bieling, and T.~Surrey, 2009.
\newblock {Obstacles on the Microtubule Reduce the Processivity of Kinesin-1 in
  a Minimal In Vitro System and in Cell Extract}.
\newblock \emph{Biophysical Journal} 96:3341--3353.

\bibitem[Seitz and Surrey(2006)]{Seitz2006a}
Seitz, A., and T.~Surrey, 2006.
\newblock {Processive movement of single kinesins on crowded microtubules
  visualized using quantum dots}.
\newblock \emph{The EMBO Journal} 25:267--277.

\bibitem[Berliner et~al.(1995)Berliner, Young, Anderson, Mahtani, and
  Gelles]{Berliner1995}
Berliner, E., E.~C. Young, K.~Anderson, H.~K. Mahtani, and J.~Gelles, 1995.
\newblock {Failure of a single-headed kinesin to track parallel to microtubule
  protofilaments}.
\newblock \emph{Nature} 373:718--721.

\bibitem[Lipowsky et~al.(2001)Lipowsky, Klumpp, and
  Nieuwenhuizen]{Lipowsky2001}
Lipowsky, R., S.~Klumpp, and T.~M. Nieuwenhuizen, 2001.
\newblock {Random Walks of Cytoskeletal Motors in Open and Closed
  Compartments}.
\newblock \emph{Physical Review Letters} 87:108101.

\bibitem[Parmeggiani et~al.(2003)Parmeggiani, Franosch, and
  Frey]{Parmeggiani2003}
Parmeggiani, A., T.~Franosch, and E.~Frey, 2003.
\newblock {Phase Coexistence in Driven One-Dimensional Transport}.
\newblock \emph{Physical Review Letters} 90:086601.

\bibitem[Parmeggiani et~al.(2004)Parmeggiani, Franosch, and
  Frey]{Parmeggiani2004}
Parmeggiani, A., T.~Franosch, and E.~Frey, 2004.
\newblock {Totally asymmetric simple exclusion process with Langmuir kinetics}.
\newblock \emph{Physical Review E} 70:046101.

\bibitem[Pierobon et~al.(2006)Pierobon, Frey, and Franosch]{Pierobon2006a}
Pierobon, P., E.~Frey, and T.~Franosch, 2006.
\newblock {Driven lattice gas of dimers coupled to a bulk reservoir}.
\newblock \emph{Physical Review E} 74:031920.

\bibitem[Melbinger et~al.(2011)Melbinger, Reichenbach, Franosch, and
  Frey]{Melbinger2010}
Melbinger, A., T.~Reichenbach, T.~Franosch, and E.~Frey, 2011.
\newblock {Driven transport on parallel lanes with particle exclusion and
  obstruction}.
\newblock \emph{Physical Review E} 83:031923.

\bibitem[Celis-Garza et~al.(2015)Celis-Garza, Teimouri, and
  Kolomeisky]{Celis-Garza2015}
Celis-Garza, D., H.~Teimouri, and A.~B. Kolomeisky, 2015.
\newblock {Correlations and symmetry of interactions influence collective
  dynamics of molecular motors}.
\newblock \emph{Journal of Statistical Mechanics: Theory and Experiment}
  2015:P04013.

\bibitem[Chandel et~al.(2015)Chandel, Chaudhuri, and Muhuri]{Chandel2015}
Chandel, S., A.~Chaudhuri, and S.~Muhuri, 2015.
\newblock {Collective transport of weakly interacting molecular motors with
  Langmuir kinetics}.
\newblock \emph{EPL (Europhysics Letters)} 110:18002.

\bibitem[Teimouri et~al.(2015)Teimouri, Kolomeisky, and
  Mehrabiani]{Teimouri2015}
Teimouri, H., A.~B. Kolomeisky, and K.~Mehrabiani, 2015.
\newblock {Theoretical analysis of dynamic processes for interacting molecular
  motors}.
\newblock \emph{Journal of Physics A: Mathematical and Theoretical} 48:065001.

\bibitem[Vuijk et~al.(2015)Vuijk, Rens, Vahabi, MacKintosh, and
  Sharma]{Vuijk2015a}
Vuijk, H.~D., R.~Rens, M.~Vahabi, F.~C. MacKintosh, and A.~Sharma, 2015.
\newblock {Driven diffusive systems with mutually interactive Langmuir
  kinetics}.
\newblock \emph{Physical Review E} 91:032143.

\bibitem[Messelink et~al.(2016)Messelink, Rens, Vahabi, MacKintosh, and
  Sharma]{Messelink2016}
Messelink, J., R.~Rens, M.~Vahabi, F.~C. MacKintosh, and A.~Sharma, 2016.
\newblock {On-site residence time in a driven diffusive system: Violation and
  recovery of a mean-field description}.
\newblock \emph{Physical Review E} 93:012119.

\bibitem[Gupta(2016)]{Gupta2015}
Gupta, A.~K., 2016.
\newblock {Collective Dynamics on a Two-Lane Asymmetrically Coupled TASEP with
  Mutually Interactive Langmuir Kinetics}.
\newblock \emph{Journal of Statistical Physics} 162:1571--1586.

\bibitem[Gillespie(1977)]{Gillespie1977a}
Gillespie, D.~T., 1977.
\newblock {Exact stochastic simulation of coupled chemical reactions}.
\newblock \emph{The Journal of Physical Chemistry} 81:2340--2361.

\bibitem[Nogales et~al.(1998)Nogales, Wolf, and Downing]{Nogales1998}
Nogales, E., S.~G. Wolf, and K.~H. Downing, 1998.
\newblock {Structure of the $\alpha$$\beta$ tubulin dimer by electron
  crystallography}.
\newblock \emph{Nature} 391:199--203.

\bibitem[Yildiz(2004)]{Yildiz2004}
Yildiz, A., 2004.
\newblock {Kinesin Walks Hand-Over-Hand}.
\newblock \emph{Science} 303:676--678.

\bibitem[Hua et~al.(1997)Hua, Young, Fleming, and Gelles]{Hua1997}
Hua, W., E.~C. Young, M.~L. Fleming, and J.~Gelles, 1997.
\newblock {Coupling of kinesin steps to ATP hydrolysis}.
\newblock \emph{Nature} 388:390--393.

\bibitem[Vale et~al.(1996)Vale, Funatsu, Pierce, Romberg, Harada, and
  Yanagida]{Vale1996}
Vale, R.~D., T.~Funatsu, D.~W. Pierce, L.~Romberg, Y.~Harada, and T.~Yanagida,
  1996.
\newblock {Direct observation of single kinesin molecules moving along
  microtubules}.
\newblock \emph{Nature} 380:451--453.

\bibitem[Howard and Hyman(2003)]{Howard2003}
Howard, J., and A.~a. Hyman, 2003.
\newblock {Dynamics and mechanics of the microtubule plus end}.
\newblock \emph{Nature} 422:753--758.

\bibitem[Ray(1993)]{Ray1993}
Ray, S., 1993.
\newblock {Kinesin follows the microtubule's protofilament axis}.
\newblock \emph{The Journal of Cell Biology} 121:1083--1093.

\bibitem[Howard(1996)]{Howard1996}
Howard, J., 1996.
\newblock {The Movement of Kinesin Along Microtubules}.
\newblock \emph{Annual Review of Physiology} 58:703--729.

\bibitem[Hyman(1995)]{Hyman1995}
Hyman, A.~A., 1995.
\newblock {Structural changes accompanying GTP hydrolysis in microtubules:
  information from a slowly hydrolyzable analogue guanylyl-(alpha,beta)-
  methylene-diphosphonate}.
\newblock \emph{The Journal of Cell Biology} 128:117--125.

\bibitem[Schneider et~al.(2015)Schneider, Korten, Walter, and
  Diez]{Schneider2015}
Schneider, R., T.~Korten, W.~J. Walter, and S.~Diez, 2015.
\newblock {Kinesin-1 Motors Can Circumvent Permanent Roadblocks by
  Side-Shifting to Neighboring Protofilaments}.
\newblock \emph{Biophysical Journal} 108:2249--2257.

\bibitem[Jeune-Smith and Hess(2010)]{Jeune-Smith2010}
Jeune-Smith, Y., and H.~Hess, 2010.
\newblock {Engineering the length distribution of microtubules polymerized in
  vitro}.
\newblock \emph{Soft Matter} 6:1778.

\bibitem[Nagel and Schreckenberg(1992)]{Nagel1992}
Nagel, K., and M.~Schreckenberg, 1992.
\newblock {A cellular automaton model for freeway traffic}.
\newblock \emph{Journal de Physique I} 2:2221--2229.

\bibitem[Weber and Frey(2017)]{Weber2016}
Weber, M.~F., and E.~Frey, 2017.
\newblock {Master equations and the theory of stochastic path integrals}.
\newblock \emph{Reports on Progress in Physics} 80:046601.

\bibitem[Lakatos and Chou(2003)]{Lakatos2003a}
Lakatos, G., and T.~Chou, 2003.
\newblock {Totally asymmetric exclusion processes with particles of arbitrary
  size}.
\newblock \emph{Journal of Physics A: Mathematical and General} 36:2027--2041.

\bibitem[Shaw et~al.(2003)Shaw, Zia, and Lee]{Shaw2003}
Shaw, L.~B., R.~K.~P. Zia, and K.~H. Lee, 2003.
\newblock {Totally asymmetric exclusion process with extended objects: A model
  for protein synthesis}.
\newblock \emph{Physical Review E} 68:021910.

\bibitem[Block et~al.(2003)Block, Asbury, Shaevitz, and Lang]{Block2003a}
Block, S.~M., C.~L. Asbury, J.~W. Shaevitz, and M.~J. Lang, 2003.
\newblock {Probing the kinesin reaction cycle with a 2D optical force clamp}.
\newblock \emph{Proceedings of the National Academy of Sciences}
  100:2351--2356.

\bibitem[Guydosh and Block(2009)]{Guydosh2009}
Guydosh, N.~R., and S.~M. Block, 2009.
\newblock {Direct observation of the binding state of the kinesin head to the
  microtubule}.
\newblock \emph{Nature} 461:125--128.

\bibitem[Mickolajczyk et~al.(2015)Mickolajczyk, Deffenbaugh, {Ortega Arroyo},
  Andrecka, Kukura, and Hancock]{Mickolajczyk2015}
Mickolajczyk, K.~J., N.~C. Deffenbaugh, J.~{Ortega Arroyo}, J.~Andrecka,
  P.~Kukura, and W.~O. Hancock, 2015.
\newblock {Kinetics of nucleotide-dependent structural transitions in the
  kinesin-1 hydrolysis cycle}.
\newblock \emph{Proceedings of the National Academy of Sciences}
  112:E7186--E7193.

\bibitem[Moyer et~al.(1998)Moyer, Gilbert, and Johnson]{Moyer1998}
Moyer, M.~L., S.~P. Gilbert, and K.~A. Johnson, 1998.
\newblock {Pathway of ATP Hydrolysis by Monomeric and Dimeric Kinesin}.
\newblock \emph{Biochemistry} 37:800--813.

\bibitem[Hancock and Howard(1999)]{Hancock1999}
Hancock, W.~O., and J.~Howard, 1999.
\newblock {Kinesin's processivity results from mechanical and chemical
  coordination between the ATP hydrolysis cycles of the two motor domains}.
\newblock \emph{Proceedings of the National Academy of Sciences}
  96:13147--13152.

\bibitem[Milic et~al.(2014)Milic, Andreasson, Hancock, and Block]{Milic2014}
Milic, B., J.~O.~L. Andreasson, W.~O. Hancock, and S.~M. Block, 2014.
\newblock {Kinesin processivity is gated by phosphate release}.
\newblock \emph{Proceedings of the National Academy of Sciences}
  111:14136--14140.

\bibitem[Clancy et~al.(2011)Clancy, Behnke-Parks, Andreasson, Rosenfeld, and
  Block]{Clancy2011}
Clancy, B.~E., W.~M. Behnke-Parks, J.~O.~L. Andreasson, S.~S. Rosenfeld, and
  S.~M. Block, 2011.
\newblock {A universal pathway for kinesin stepping}.
\newblock \emph{Nature Structural {\&} Molecular Biology} 18:1020--1027.

\bibitem[Hancock(2016)]{Hancock2016}
Hancock, W.~O., 2016.
\newblock {The Kinesin-1 Chemomechanical Cycle: Stepping Toward a Consensus}.
\newblock \emph{Biophysical Journal} 110:1216--1225.

\bibitem[Andreasson et~al.(2015)Andreasson, Milic, Chen, Guydosh, Hancock, and
  Block]{Andreasson2015}
Andreasson, J.~O., B.~Milic, G.-Y. Chen, N.~R. Guydosh, W.~O. Hancock, and
  S.~M. Block, 2015.
\newblock {Examining kinesin processivity within a general gating framework}.
\newblock \emph{eLife} 4:1--44.

\bibitem[Isojima et~al.(2016)Isojima, Iino, Niitani, Noji, and
  Tomishige]{Isojima2016}
Isojima, H., R.~Iino, Y.~Niitani, H.~Noji, and M.~Tomishige, 2016.
\newblock {Direct observation of intermediate states during the stepping motion
  of kinesin-1}.
\newblock \emph{Nature Chemical Biology} 12:290--297.

\bibitem[Verbrugge et~al.(2009)Verbrugge, van~den Wildenberg, and
  Peterman]{Verbrugge2009}
Verbrugge, S., S.~M. van~den Wildenberg, and E.~J. Peterman, 2009.
\newblock {Novel Ways to Determine Kinesin-1's Run Length and Randomness Using
  Fluorescence Microscopy}.
\newblock \emph{Biophysical Journal} 97:2287--2294.

\bibitem[Khataee et~al.(2018)Khataee, Naseri, Zhong, and Liew]{Khataee2017}
Khataee, H., S.~Naseri, Y.~Zhong, and A.~W.-C. Liew, 2018.
\newblock {Unbinding of Kinesin from Microtubule in the Strongly Bound States
  Enhances under Assisting Forces}.
\newblock \emph{Molecular Informatics} 37:1700092.

\bibitem[Soza{\'{n}}ski et~al.(2015)Soza{\'{n}}ski, Ruhnow, Wi{\'{s}}niewska,
  Tabaka, Diez, and Ho{\l}yst]{Sozanski2015}
Soza{\'{n}}ski, K., F.~Ruhnow, A.~Wi{\'{s}}niewska, M.~Tabaka, S.~Diez, and
  R.~Ho{\l}yst, 2015.
\newblock {Small Crowders Slow Down Kinesin-1 Stepping by Hindering Motor
  Domain Diffusion}.
\newblock \emph{Physical Review Letters} 115:218102.

\end{thebibliography}

}

\end{document}